# REAL TIME COLLABORATIVE PLATFORM FOR LEARNING AND TEACHING FOREIGN LANGUAGES

*Ilya V. Osipov, Anna Y. Prasikova*

*i2istudy SIA, Krišjāņa Barona Iela, 130 k-10, Rīga, Lv-1012,Latvija, ilya@osipov.ru*

*Alex A. Volinsky*

*Department of Mechanical Engineering University of South Florida, E. Fowler Ave., ENB118, Tampa FL 33620, USA, volinsky@usf.edu*

*CORRESPONDING AUTHORS:*

*Ilya V. Osipov, Alex A. Volinsky*

## Abstract

The paper describes a novel social network-based open educational resource for learning foreign languages in real time from native speakers, based on the predefined teaching materials. This virtual learning platform, named i2istudy, eliminates misunderstanding by providing prepared and predefined scenarios, enabling the participants to understand each other and, as a consequence, to communicate freely. The system allows communication through the real time video and audio feed. In addition to establishing the communication, it tracks the student progress and allows rating the instructor, based on the learner's experience. The system went live in April 2014, and had over six thousand active daily users, with over 40,000 total registered users. Currently monetization is being added to the system, and time will show how popular the system will become in the future.

## Introduction

Open educational resources (OER) have recently become quite popular in the area of computer assisted language learning (Cushion, 2005, Coryell and Chlup, 2007, Herasim, 2012, Thomas & Evans, 2014). Currently there are several educational services on the market with a considerable amount of OERs that provide an opportunity to learn foreign languages (livemocha.com (Sevilla-Paóvn, Martínez-Sáez, Gimeno Sanz and Seiz-Ortiz, 2012), www.learn-english-online.org (Giles, 2012), www.duolingo.com (Rutkin, 2014)). Most of these systems are automated, i.e. don't require live human interaction. These systems can be divided into two categories: autonomous and social. Autonomous methods offer tasks, which are checked or monitored in accordance with the algorithms set up within the system (tests, quizzes, etc., Son, 2007). Social methods allow direct or indirect interaction with real people, including communication, checking assignments, etc.

(Yousefi, 2014, www.facebook.com). Such systems have been used in language learning (Wang and Chen, 2009, Donmus, 2010, Kurata, 2010, Aydin, 2014 and Toetenel, 2014). Murday, Ushida, and Chenoweth 2008, Tal and Yelenevskaya 2012, C. Cheong, V. Bruno, 2012) also attempted to integrate computer-assisted language learning systems into the educational process. Golonka, Bowles, Frank, Richardson and Freynik, 2014 wrote an excellent review on the subject. One of the currently largest and most popular systems is livemocha.com, which is a social network for the learners of foreign languages. The service started in 2007 and currently has over 14 million users. The site uses community learning approach, where students learn through live human interactions using video chat. Currently 38 languages are offered, including four levels for each language. Another system, learn-english-online.org has predefined learning materials, but it is complicated to use. There is a chat feature; however, it is not clear how to connect the users through Skype (Budiman, 2013, Hashemi and Azizinezhad, 2011). Another system, duolingo.com uses predefined materials without live human interaction, which is quite important in foreign language learning and provides better learning outcomes (Kötter, 2010, Blum-Kulka & Dvir-Gvirsman, 2010, Kim, 2014).

Currently the most popular distance language learning tool is Skype (Rao, Angelov and Nov, 2006), the real time audio-video communication system, which is not designed for this purpose (Hashemi and Azizinezhad, 2011). Thousands of small companies and individuals offer foreign language learning through Skype. Skype is even being positioned as an instructional resource (Kiziltan, 2012). Searching for "English via Skype" in Google gives over 43 million results, and similar numbers appear if the search is conducted in Spanish, Russian and other languages. There is a large demand for online foreign language education (Kozar and Sweller, 2014) However, Skype does not allow finding the person willing to teach/learn foreign languages, it does not provide learning aid materials, does not track the spent time in the user account. Several papers discuss Skype as the language learning tool (Rao, Angelov and Nov, 2006, Kiziltan, 2012, Budiman, 2013, Hashemi and Azizinezhad, 2011).

The novel method of online collaborative learning (Lee et al., 2014) of foreign languages takes social learning to a new level, using direct interaction between users, one of which has the competence as a native speaker, and the other one is the recipient of knowledge (the language learner). This virtual learning platform allows to eliminate the problem of misunderstanding each other between all languages, by providing participants with prepared and predetermined scenarios, which enables the participants to understand each other and, as a consequence, to communicate freely. This communication is informal, resulting in better learning outcomes (Lai et al., 2013). The service is utilizing task-oriented interactions (Gánem-Gutiérrez, 2009, Delahunty, Verenikina & Jones, 2013). It allows overcoming the learner's anxiety, similar to the initial attempts of Yen, Hou & Chang (2013), but on a much larger scale, currently utilizing four languages.

This system provides an opportunity for the participants to have real time audio and video communication with each other, based on the WebRTC and Adobe Air technologies, followed by interactive prompts, which are controlled by the participants in real time. The prompts consist of images, texts and videos (Schworm and Bolzer, 2014) displayed to each participant in their native understandable language. These technological solutions allow using this system with the help of virtually any personal computer with the microphone and the camera connected to the network at the same time without additional specific technical requirements. This methodology can be used to learn foreign languages from the native speakers in real time, based on the predetermined teaching materials, which simulate real life situations using task-oriented interactions (Hauck and Young, 2008, Wang, Chen and Levy, 2010, Osipov, 2013). It is also a great research tool in the area of computer-assisted learning.

# Research objectives

The i2istudy.com is currently a free multilingual web service for studying foreign languages online. The main idea of the service is based on the "time banking" principle (Válek and Jašíková, 2013, Seyfang and Longhurst, 2013). For every minute that a person teaches in the mother tongue, he/she is rewarded with a minute that can be used to learn a foreign language. This currently allows using the system free of charge (Seufert, 2014).

The name "i2istudy" comes from the idea of the "eye to eye" learning, based on the peer2peer principle (Hsu, Jub, Yen and Chang, 2007). A new model of studying with the native speaker according to a set of interactive courses was created. Every "lesson" is based on the split screen platform (Figure 1). On one half of the screen the native language teacher is shown, and on the other half is a set of audio and video slides (Figure 1). Together with the native speaker, the learner goes through the predefined content. Therefore, there is no need for the professional teacher, and theoretically every person can teach their mother tongue within this format. Furthermore, the i2istudy is a self-regulated system, as it allows every user to pick their own paste, time, level, gender and plenty of other characteristics of their instructor and the lesson, thus everyone can find their "right" teacher.

The i2istudy is an extremely social and informal approach to learning languages (Lai et al., 2013). It is not only a mere learning platform, but also a learning community that brings users together and builds relationships (Cohen, 2014). Learning foreign languages is a universal international factor uniting like-minded individuals all over the World. While studying, the system users enter into an intense communication process with a native speaker, not only studying the language, but also the culture, behavior and the manners. They not only discuss the set topics, but can also enter into a more personal communication. At the present time the system supports four languages: English, Spanish, Russian and German, however more languages can be added without significant technical changes of the system (Osipov, Volinsky, Grishin, 2014).

This virtual learning platform provides users with the opportunity to discuss particular subjects in situations where one of the users is a native speaker and an expert in the particular knowledge and the other party is the recipient of this knowledge. Each user is provided accompanying materials in accordance with the level of competence in the subject. Communication in different languages between users is made possible with the help of predefined communication scenarios for the purpose of studying a foreign language.

This capacity is provided through the combination of three components for the user interaction with each other: one component provides real time audio-visual communication for users, the next component displays text and a scenario of topics in stages used for communication in foreign languages understood by the users, and the third component allows communicating by sending and receiving instant text messages (Figure 2). The system user interface is shown in Figure 1 and schematically in Figure 2. The system automatically tracks the time, which is reported in user accounts for the purpose of time banking (Marks, 2012). The user interface, user interaction and the script are presented by means of individual cards (slides), connected by a mutual discussion topic. Each slide consists of a separate text, graphics and video in an interface, which is clear and understandable to each user, in their own native language (Figure 2). Each slide consists of a set of common fields provided in multiple languages, or in each individual language (Figure 3).

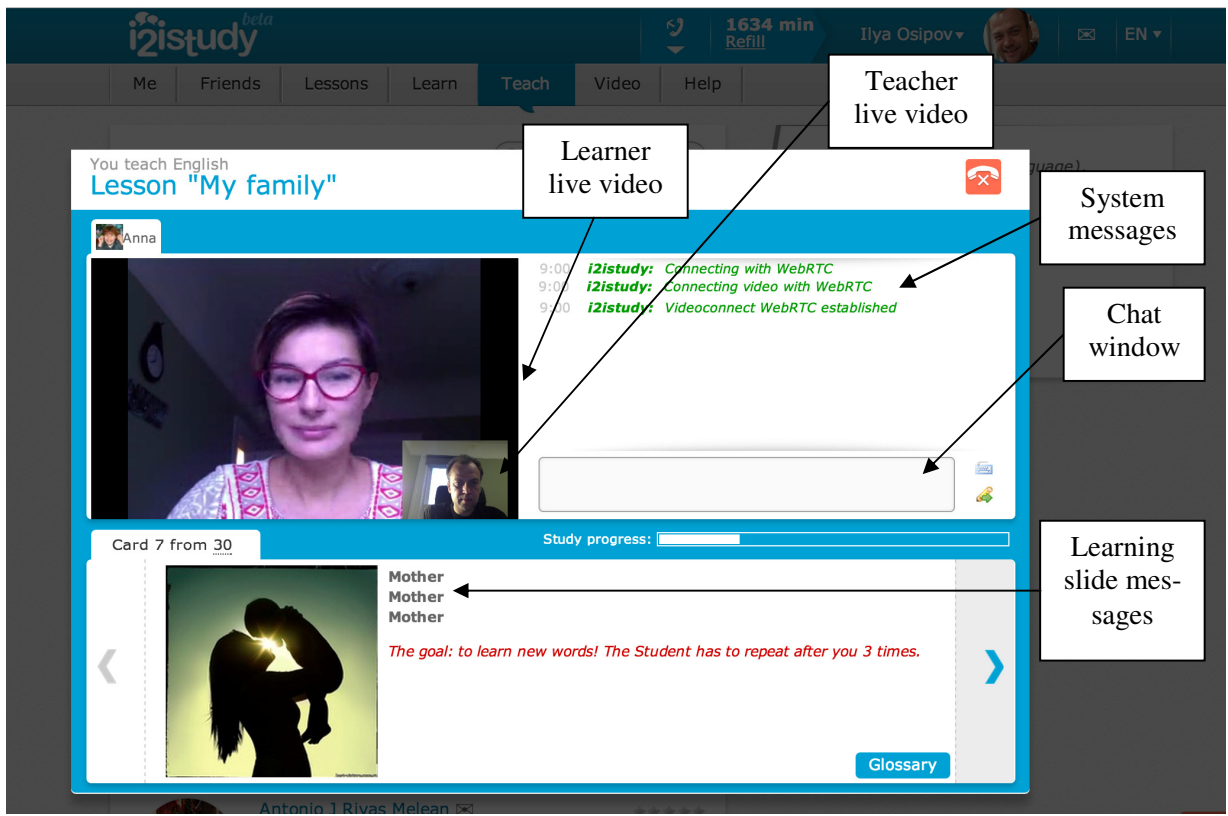

*Figure 1*. **i2istudy lesson user interface.**

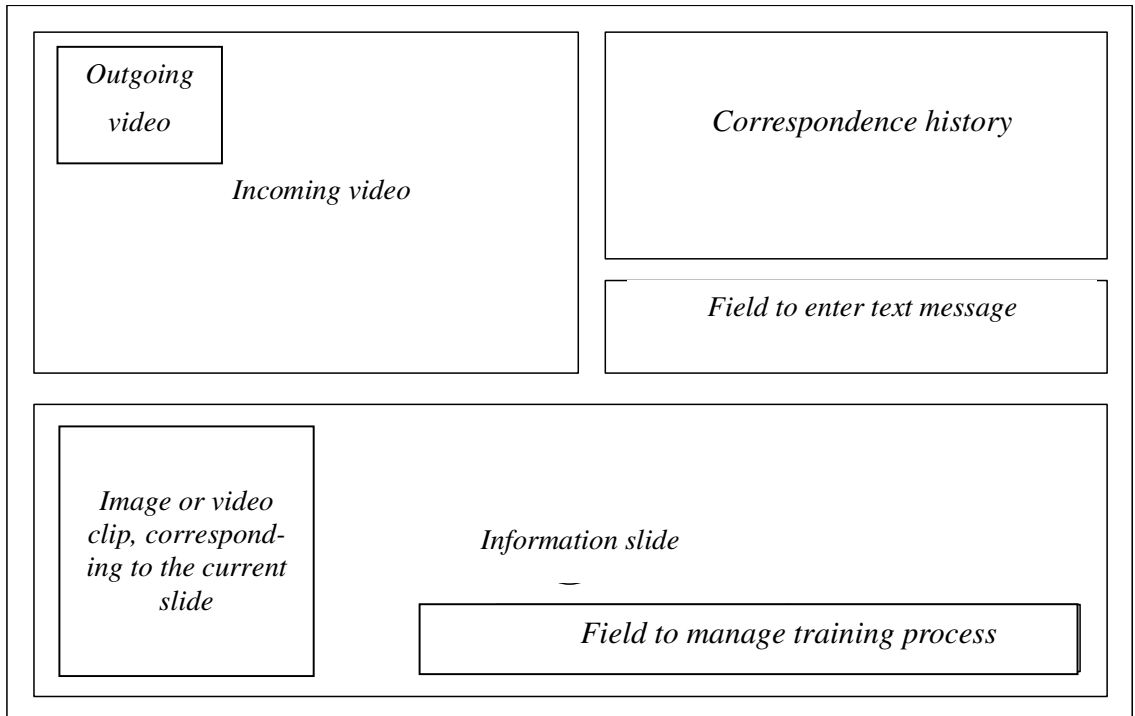

*Figure 2*. **System user interface layout schematics.**

The instructor, while being the one responsible to issue comments, dictates to the student what and how to speak, in addition to how to respond to his/her statements. The student is displayed with the minimal information required, since his/her task is to try to understand the native speaker with the amount of information provided. In general, it is assumed that the instructor is available to lead the teaching process, namely, start the training, switch the information slides and decide when to stop the lesson, as well as activate the linguistic tips for the student when necessary (Figure 4). The student also has the ability to activate available linguistic tips. For the purpose of studying foreign languages, the instructor switches the slides and reads the tasks, while the student listens and sees the native speaker, receiving minimal amount of information required for basic understanding. If the student does not understand the teacher, the text can be displayed corresponding to what the teacher said, along with its translation. Depending on the user's role (teacher/student), the slides form a coherent set of readily available information for both parties.

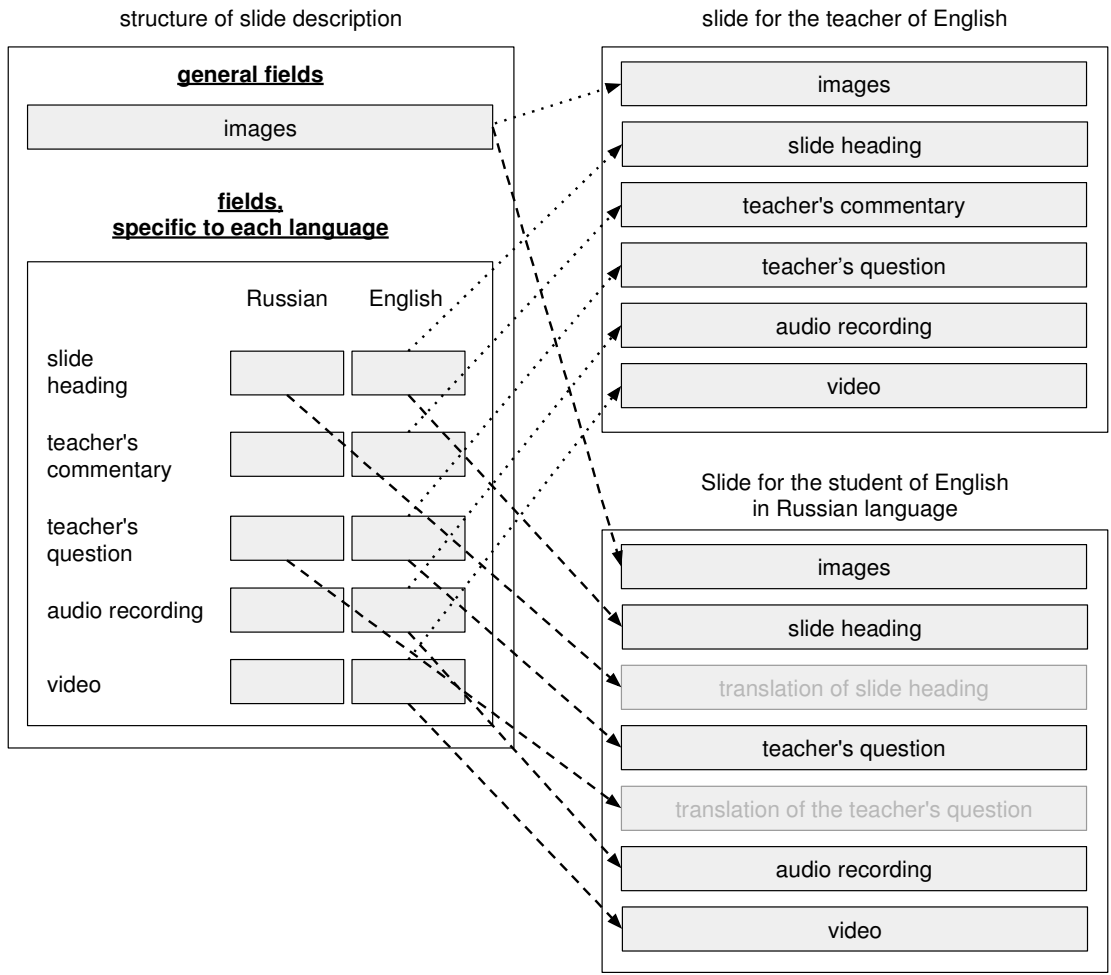

*Figure 3.* **The slide structure, description and the format, based on the different user roles.**

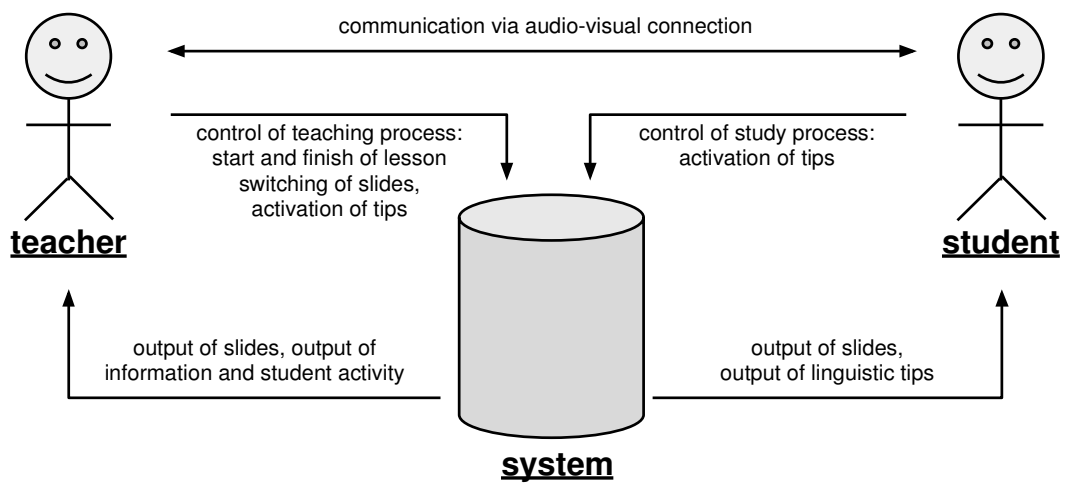

*Figure 4.* **The teacher-student interaction.**

In order to implement more complex forms of interaction, it may be necessary to have more roles. The system also allows for the connection of multiple users with individual roles in the educational process. The logic of assigning roles within the framework of the system is that each user separately determines their own solutions. For example, to organize teaching of several students by a specialist in a particular knowledge, along with the monitoring of the teaching process, it may be necessary to have the roles of a teacher, a student and a supervising controller. The teacher conveys the material to one or more students. However, the decision concerning the successful presentation and mastering of the material, for example, whether to jump to the next course or not, is taken by the controller (Figure 5). The controller oversees the educational process. For example, decisions whether the students can independently take the interactive tests available in the system, along with assessing the competency level of the teacher are taken by the controller. Currently, for the study of foreign languages in the form of a game, the system allows only two roles: the teacher and the student. However, there was a logistic problem of how the students and the teachers would be connected, thus a special module was designed, called the connector module.

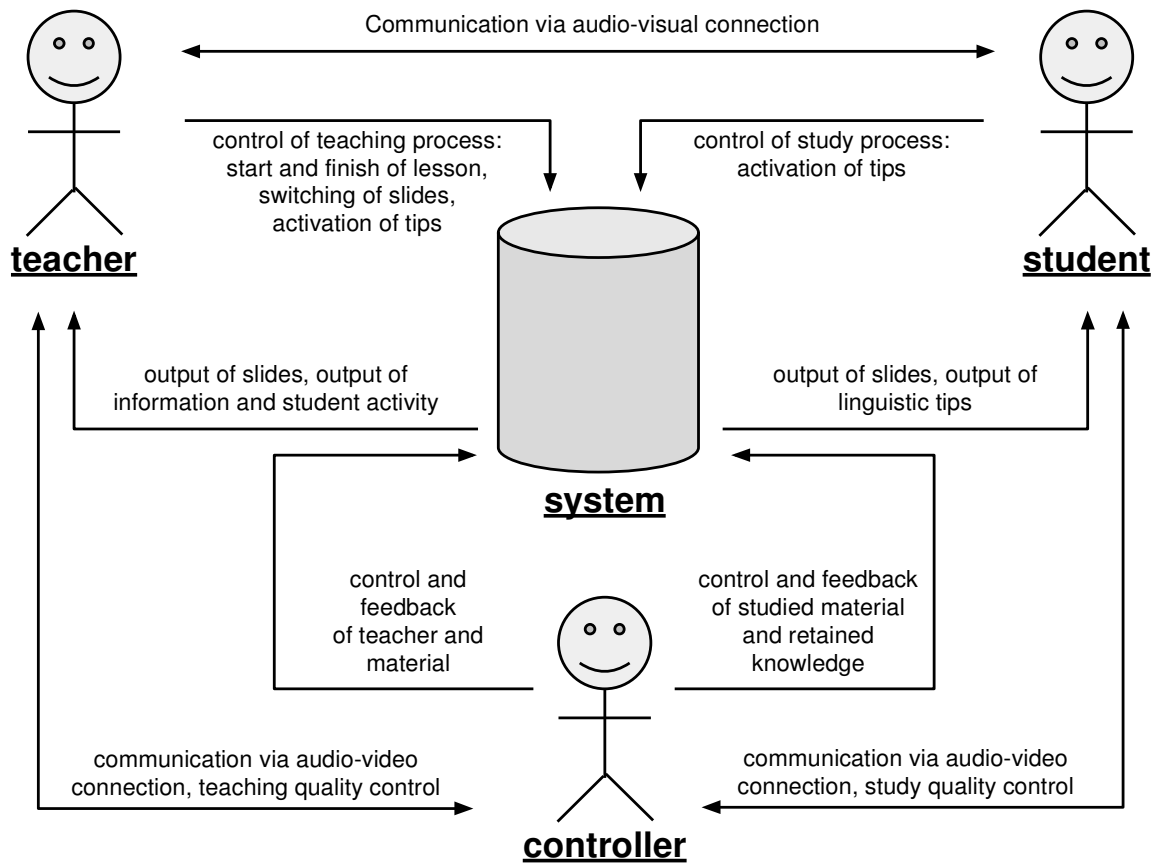

*Figure 5*. **Teacher-student-controller interaction.**

## *The connector module*

The connector functions are summarized in Figure 6. The main obstacle to the users' interaction in real-time is the search for available users in order to begin the interaction at a specific time. While trying to establish a connection with a specific user, which for some reason does not answer an incoming call, other users are deemed unavailable for the call. This is a classic problem of a phone call: when you call one number, there is no opportunity to call other recipients, since your line is busy, in addition to the fact that the current recipient of the call is also unavailable to other users.

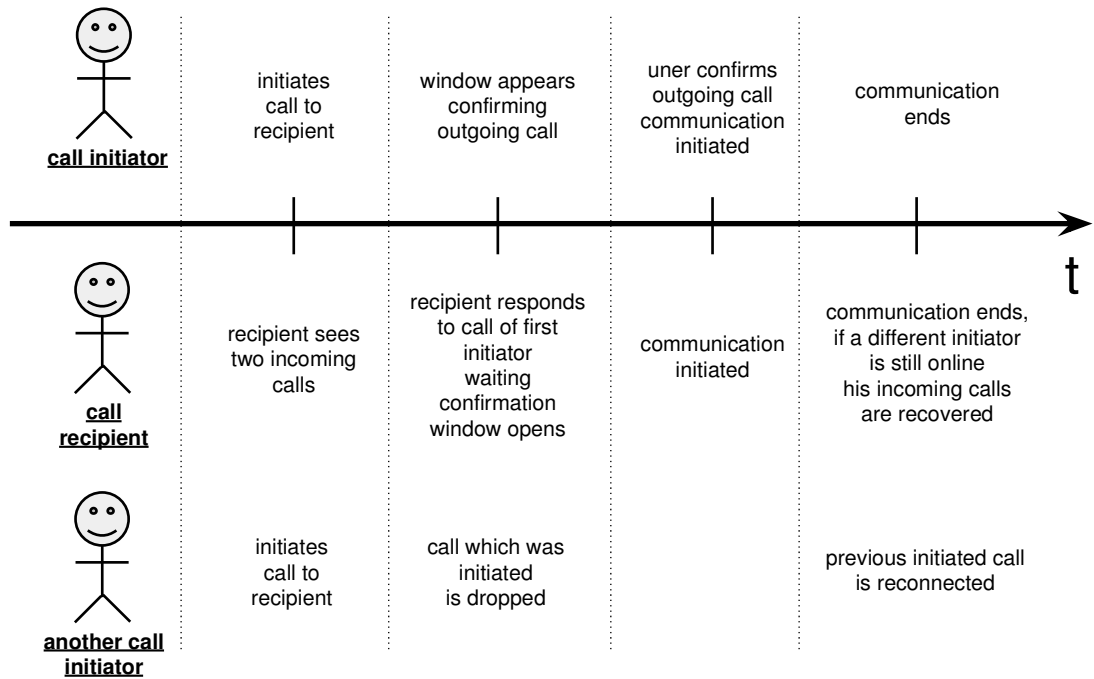

*Figure 6.* **Connector function diagram**.

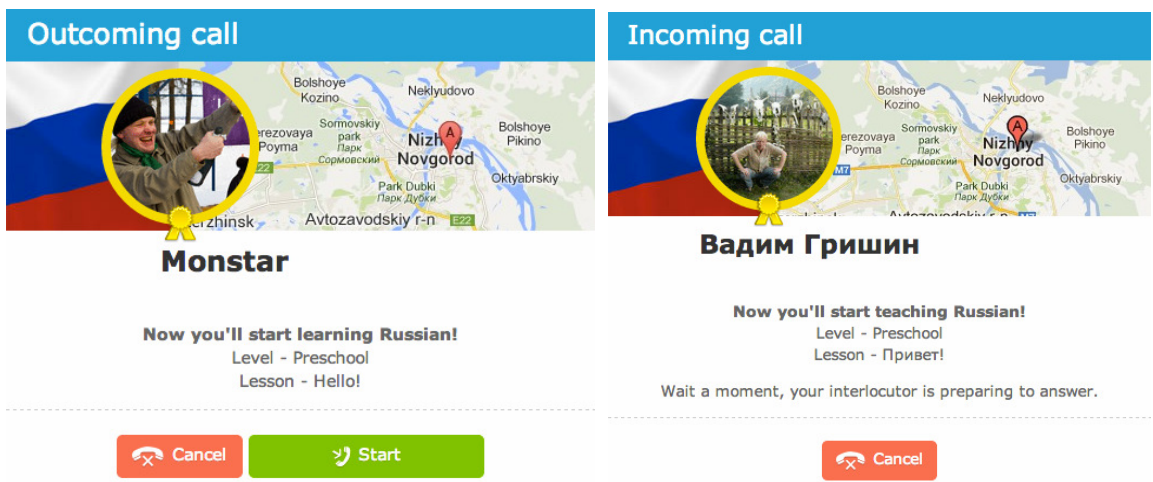

*Figure 7.* **Answering an incoming call: waiting for the confirmation of the recipient (left) and the initiator of the call confirmation (right)**.

This problem is solved by implementing the multi-call system, allowing one user to send requests to multiple recipients, increasing the response probability. This solution is not burdensome for users, as it allows monitoring the availability of users at the moment, while making the user inaccessible for additional calls. After the line is freed up, the user is given an option to connect missed calls from others who called while the line was busy.

When the user clicks on the notification for the incoming calls, both users receive a notification at the beginning of the call (Figure 7). At the same time both incoming and outgoing calls of other users will go on hold. The user's data and their incoming and outgoing call information become unavailable to all other users. Thus, for the user who answered the call, a notification is sent about the beginning of communication. In other windows and client pages, a notification is displayed that the user is active. The user receiving the call, along with the caller, both receive a notice for the administration of a connection or its cancellation. In order to initiate the connection, the user confirms that he/she is ready; thus communication is established between the clients, generating a response and confirmation.

**Software implementation and initial challenges**

The server portion was written using the PHP programming language combined with the MySQL database. The user client needs an HTML browser with the Java Script installed. Adobe Flash Air or WebRTC are used for the audio-visual real time connection (Karopoulos, Mori and Martinelli, 2013).

Initially there were multiple compatibility problems with Adobe Flash Air and WebRTC due to a large number of different browsers and their versions installed by the users. While the initial testing went quite well, there were browser combinations that could not establish a connection. The number of browser combinations was quite high. The problem was solved by collecting the statistics with further fine tuning of the browser settings and selecting the proper technology supported by the system.

Additionally, there were multiple problems associated with the new users, who didn't understanding how to use the system. The whole concept of the i2istudy.com is quite novel, and initially puzzled the newly registered users. The problem was solved by adding numerous prompts, helping the users navigate through the system. Detailed help was also developed, including frequently asked questions and videos of the demo lessons. The videos of the demo lessons were dubbed in the four languages, currently available in the system (English, Russia, German and Spanish). The users are more likely to follow the steps outlined in the videos, and the amount of questions and misunderstanding was reduced dramatically.

While currently the system requires the use of a personal computer, in the future there could be an option to use it on mobile devices as well, as mobility is the key for the language learning (Virvou, Alepis and Troussas, 2011)..

# Gamefication technology and virality

Currently the i2istudy system allows:

Choosing the teacher in the system, based on the offered language, country, gender and age. Sending requests for learning using the Connector module.

Choosing lessons from the library of themes and the language proficiency level.

Tracking the time in minutes spent teaching and learning in the individual user accounts for the purpose of time banking.

Facilitating the teaching/learning process, consisting of the three components: live video, step-by-step learning with the aid of predefined materials in the native language, and the chat window.

Besides these features, other aspects of gamefication and virality are being addressed (Fields and Cotton, 2014). Gamefincation is needed to keep the user short term and long term attention, making sure that he/she will come back and continue to use the system in the future. The following attributes are utilized. Numerical value represents the user involvement, along with the leader board. Each user activity is tracked over a month timeframe, allowing the user to compare his/her results with others, based on their respective progress. The rating system for each user allows rating learning/teaching partners after each lesson. The scores are accumulated and are visible to other users, encouraging them to improve their scores. Accessing parts of the lessons is allowed. Passing each set of lessons allows opening new lessons, similar to a computer game, intriguing and entertaining the user. Each lesson progress is tracked with the progress bar, showing how many slides have been studied, and how many slides are still left in the lesson. Information about the newly learned words and phrases is also displayed. Custom decals and badges are awarded to the users, based on their activity in the system. After successfully completing several lessons, the user gets the "Expert" status, which is promoted as an achievement, and is visible to other users. Information about completed lessons can be readily shared in social networks as well. The user gets more involved in the game, achieving higher status with more activity in the system.

Besides keeping the current users involved, the system benefits from attracting the new users, using the following virality mechanisms:

Inviting friends through social networks, like Facebook and others.

Encouraging users to send invitations to his/her friends to attract them to the system. Currently each user gets extra 30 minutes of free lessons for each invited friend, who joins.

Sharing learning/teaching results in social networks if desired.

Detailed description of the system gamification, virality and user retention is a subject of a separate publication. Based in the conducted market analysis, the users demanded system monetization, which is currently being implemented.

## *Conclusions*

The paper demonstrates the novel approach to learning foreign languages online from the native speakers. The system went live in April 2014, and had over 6,000 active daily users, with over 40,000 registered users. Over 300 new users registered daily. There were over 10 hours of live learning and teaching per day and 12 hours spent with an automated teacher (recorded lessons). These numbers have been growing even without active advertising. However, the project has been closed for three months to add monetization and other system improvements. The developers hope that the i2istudy.com open educational resource will become quite popular with the learners of foreign languages around the World.

Golonka, E.M., Bowles, A.R., Frank V.M., Richardson, D.L., Freynik, S. (2014). Technologies for foreign language learning: a review of technology types and their effectiveness. *Computer Assisted Language Learning*, 27(1), 70-105.

Harasim, L. (2012). Learning theory and online technologies, *Routledge, New York and London*, 191, ISBN 978-0-41-599975-5

Hashemi, M., Azizinezhad, M. (2011). The capabilities of Oovoo and Skype for language education. *Procedia Social and Behavioral Sciences*, 28, 50.

Hauck, M., Young, B.L. (2008). Telecollaboration in multimodal environments: the impact on task design and learner interaction. *Computer Assisted Language Learning*, 21(2), 87-124.

Hsu, M.-H., Jub, T.L., Yen, C.H., Chang, C.-M. (2007). Knowledge sharing behavior in virtual communities: The relationship between trust, self-efficacy, and outcome expectations. *International Journal of Human-Computer Studies*, 65(2), 153-169.

Karopoulos, G., Mori, P., Martinelli, F. (2013). Usage control in SIP-based multimedia delivery. *Computers & Security*, 39(B), 406.

Kim, H.Y. (2014). Learning opportunities in synchronous computer-mediated communication and face-to-face interaction. *Computer Assisted Language Learning*, 27(1), 26-43.

Kiziltan, N. (2012). Teaching Turkish through Teletandem. *Procedia Social and Behavioral Sciences*, 46, 3363.

Kozar, O., Sweller, N. (2014). An exploratory study of demographics, goals and expectations of private online language learners in Russia. *System*, 45, 39.

Kötter, M. (2010). MOOrituri te salutant? Language learning through MOO-based synchronous exchanges between learner tandems. *Computer Assisted Language Learning*, 14(3-4), 289-304.

Kurata, N. (2010). Opportunities for foreign language learning and use within a learner's informal social networks. *Mind, Culture, and Activity*, 17, 382.

Lai, K.-W., Khaddage, F. , Knezek, G. (2013). Blending student technology experiences in formal and informal learning. *Journal of Computer Assisted Learning*, 29(5), 414-425.
12